\documentclass[conference]{IEEEtran}
\usepackage{cite}

\ifCLASSINFOpdf
\else
\fi
\usepackage{amsmath}
\usepackage[linesnumbered,ruled,vlined]{algorithm2e}
\usepackage{algorithmicx}

\usepackage{pifont}

\usepackage{pgfplots}
\usepackage{tikz}
\usepackage{graphicx}
\usepackage{url}

\IEEEoverridecommandlockouts
\begin{document}
%
\title{Caching Techniques to Improve Latency in Serverless Architectures}


\author{\IEEEauthorblockN{Bishakh Chandra Ghosh\IEEEauthorrefmark{4}, Sourav Kanti Addya\IEEEauthorrefmark{4}\thanks{\IEEEauthorrefmark{4} Equal contribution}, Nishant Baranwal Somy, Shubha Brata Nath, \\Sandip Chakraborty, and Soumya K Ghosh}
	\IEEEauthorblockA{Department of Computer Science and Engineering\\
		Indian Institute of Technology Kharagpur, India
				\\Email: \{ghoshbishakh, kanti.sourav, somy1997, nath.shubha\}@gmail.com, \{sandipc, skg\}@cse.iitkgp.ac.in}
}

\maketitle

\begin{abstract}
Serverless computing has gained a significant traction in recent times because of its simplicity of development, deployment and fine-grained billing. However, while implementing complex services comprising databases, file stores, or more than
one serverless function, the performance in terms of latency of serving requests often degrades severely. In this work, we analyze different serverless architectures with AWS Lambda services and compare their performance in terms of latency with a traditional virtual machine (VM) based approach. We observe that database access latency in serverless architecture is almost 14 times than that in VM based setup. Further, we introduce some caching strategies which can improve the response time significantly, and compare their performance.
\end{abstract}


\IEEEpeerreviewmaketitle

\section{Introduction}\label{intro}
In the current age of web services, the development and deployment of web applications in the cloud is convenient than ever before with the advent of `serverless computing' \cite{Adzic:2017}. The application developers and software service providers do not need to even estimate their server specifications during the service level agreement negotiation. Moreover, they can safely leave the worries of scaling their service to the cloud service platforms. All these are made possible by the concept of serverless application or Function as a Service (FaaS) \cite{Wang_usenix_2018}. Popular cloud providers such as Amazon, Microsoft, and Google already introduced their serverless solutions as \textit{AWS Lambda}\footnote{\url{https://aws.amazon.com/lambda/}}, \textit{Azure Function}\footnote{\url{https://azure.microsoft.com/en-in/solutions/serverless/}}, and \textit{Google Cloud Function}\footnote{\url{https://cloud.google.com/functions/}} respectively. Apart from these big players, some new solutions have also started providing the FaaS service \cite{servelless_online_2019}.

\noindent Serverless platform lets application developers to focus on the core product and business logic instead of spending time on responsibilities like setting up a server, installing and updating operating system and runtime environmets, managing access control policies, provisioning, right-sizing, scaling, and availability \cite{Lloyd_2018}. By building the application on a serverless platform using FaaS, the platform manages these responsibilities for the developer. Thus, serverless functions together with complementary services like managed database services, file storage services, key-value stores, API gateways, etc., provide the following benefits over traditional cloud virtual machine (VM) based setups -- i) No server management, ii) Flexible scaling, iii) High availability, iv) No idle capacity, and v) Fine-grained billing.


Presently, maximum adopters of this technology are startups, who seek to quickly develop their services and scale their resources efficiently.
Serverless architecture is also a perfect candidate for 
a wide range of applications, ranging from a simple database application to complex data analytics pipelines \cite{Bhattacharjee_USENIX_2019}.
However, everything comes at a price. Our analysis in this work shows that by using serverless computing although we are gaining in terms of ease of use, scalability and pricing, there is a significant trade-off with the service quality, especially in terms of response time.
%
%

In the literature, a few recent works have significantly explored serverless computing from different perspectives. In \cite{Akkus_Sand_Usenix_2018}, a new serverless computing with better resource efficiency and more elasticity than the existing serverless platforms is discussed. To achieve these properties, authors have introduced a model SAND, based on an hierarchical message bus and application level sandboxing. Here, the design and implementation of SAND, as well as experience in building and deploying serverless applications on it are presented. In \cite{Oakes_USENIX_2018}, the authors have analyzed linux container primitives, identifying scalability bottlenecks related to storage and network isolation where there is a container system optimized for serverless workloads. Based on these findings, they have implemented SOCK, a container system optimized for serverless workloads model. They have identified container initialization and package dependencies as the common causes of slow Lambda startup. In \cite{Hong_USENIX_2018}, authors have advocated six design patterns using serverless concept to develop cloud security service. They explain the pros and represent the applications for each design. They have also introduced a threat intelligence platform which stores logs from different sources, alerts malicious events, initiate possible action for those activities.


Though advantageous in focusing on the user's business logic rather than the platform management, serverless architectures have latency related issues as it is stateless and need to communicate with other serverless functions and other components such as data stores. In this paper, we analyze serverless architectures concerning inter-component latency and overall response time of serving requests. First, we take the simplest use case of a database application and compare the performance with the traditional VM based approach. We observe that compared to VM based systems, serverless architecture suffers from very high latency for database access. Next, we analyze a more complex data analytics pipeline involving multiple functions and observe how the scaling up of the architecture impacts the overall latency of the service. Finally, we propose an in-memory internal caching strategy and compare it with \texttt{AWS ElastiCache}\footnote{https://aws.amazon.com/elasticache/}. From the comparison, we observe that our proposed system improves the response time significantly.

\section{Latency Analysis}\label{latencyanalysis}
In order to understand the latency of serverless architectures in \texttt{AWS Lambda} services, we consider two practical use cases:
(1) Simple database application and 
(2) Complex data analytics pipeline. We compare the latency of the serverless system with a traditional cloud VM based setup and see how this latency is affected as the system scales to more complex architectures.

\begin{figure}[ht]
	\centering
	\includegraphics[width=0.65\linewidth]{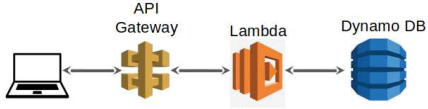}
	\caption{Serverless architecture for simple DB application}\label{fig:simpleDB_serverless}
\end{figure}

\begin{figure}[ht]
	\centering
	\includegraphics[width=0.65\linewidth]{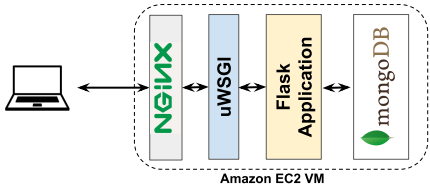}
	\caption{VM based setup using Flask and MongoDB on AWS EC2}\label{fig:vmbasedsetup}
\end{figure}

\begin{figure}[ht]
	\centering
	\includegraphics[width=0.8\linewidth]{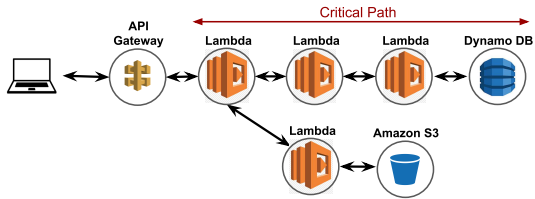}
	\caption{Serverless architecture for data analytics pipeline application}\label{fig:DataAnalytics_serverless}
\end{figure}

\subsection{Methodology}
(1) \textbf{Simple database application}: This involves only one Lambda function and one database. As the serverless functions are stateless and unlike VM, we cannot save any data on it, we use \texttt{Amazon DynamoDB}\footnote{https://aws.amazon.com/dynamodb/}, which is a key-value and document database. In order to access the service, we connect the Lambda function to an API gateway through which it can accept HTTP requests. The architecture is depicted in Figure \ref{fig:simpleDB_serverless}. We implement a simple application in a single Lambda function using \texttt{Node.js} to store and retrieve account details of users from the database. In order to compare the performance of this setup relative to the traditional cloud VM based approach, we implement a web service with the same functionality in a VM on \texttt{Amazon EC2}\footnote{https://aws.amazon.com/ec2/} service. For this setup we use \texttt{Python Flask}\footnote{http://flask.palletsprojects.com} web framework along with \texttt{MongoDB}\footnote{https://www.mongodb.com/} as a database (Figure \ref{fig:vmbasedsetup}). The Flask application is served through \texttt{Nginx}\footnote{http://nginx.org/} web server and \texttt{uWSGI}\footnote{https://uwsgi-docs.readthedocs.io/en/latest/} application server. All these components run on the same VM having one vCPU and 1 GB of memory. We deploy both the serverless and server based setups in five data centers in different regions namely, Mumbai, London, California, Canada central, and Singapore. In order to make a fair comparison, to eliminate the factor of network quality delay from the end-user, we only consider the time taken to process the user's request by the serverless setup and the traditional VM based setup. For serverless setup, we use \texttt{CloudWatch}\footnote{https://aws.amazon.com/cloudwatch/} logs, and for server based setup, we use \texttt{Nginx} logs to get the response time of the service.

(2) \textbf{Complex data analytics pipeline}: We take into account multiple Lambda functions which are common for data analytics pipelines, data mining workflows, and any other complex applications involving multiple steps and processes. Any such workflow can be depicted as a graph (Figure \ref{fig:DataAnalytics_serverless}), where the vertices are serverless components such as logic units like Lambda functions, databases like \texttt{DynamoDB}, file stores like \texttt{Amazon S3}, key-value stores such as \texttt{Redis}, etc. There exists an edge between two such vertices if one component calls the other component synchronously, and waits for its response. In other words, there is an edge between two components if one component depends on another component.
Thus, it is expected that the response time of the service will be equal to the sum of the computation time of the components in the longest path of the graph which we call the \textit{critical path}. However, since the functions and components are managed by the provider, here AWS, we do not have any control of their deployment except ensuring that they are deployed in the same region. As a result, the Lambda functions, databases, and other services are most likely deployed in different hosts and must communicate through the network which incurs some delay. This delay between each component ultimately has a severe cascading effect on the overall response time of the application. We analyze the overall latency with different workflows of varying critical path lengths involving a series of Lambda functions terminating at \texttt{DynamoDB}.

\begin{figure}[h]
	\center
	\includegraphics[width=0.9\linewidth]{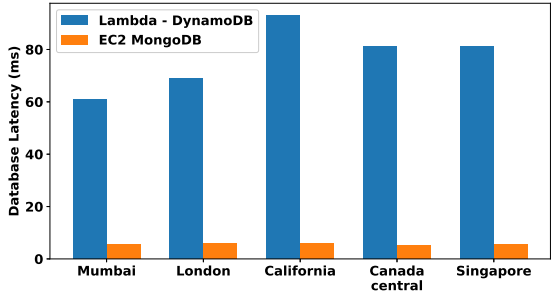}
	\caption{Database access latency in serverless and VM based architecture}\label{fig:DynamoDBvsMongoDb}
\end{figure}

\begin{figure}[h]
	\center
	\includegraphics[width=\linewidth]{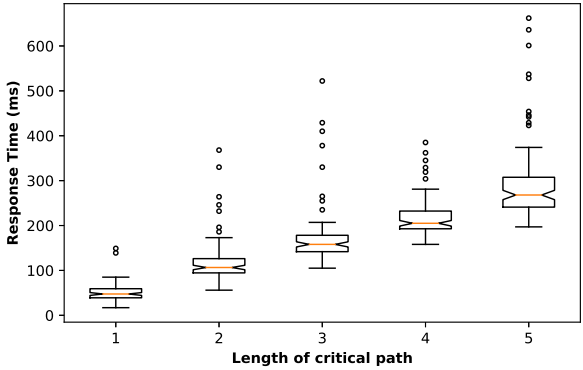}
	\caption{Latency with increasing critical path length}\label{fig:serverlessCascadingData.png}
\end{figure}

\subsection{Observations}

For the first scenario, the simple database application involving only a single database read or write operation per request, we measure the response time of the system per user request. We deploy both the serverless setup and the VM based setup in five regions. We observe that the  response time for the serverless architecture is significantly higher than the VM based setup. Our analysis shows that the reason of this higher latency is not because of higher processing time by the Lambda function but mostly due to the latency of the database access.  In Figure \ref{fig:DynamoDBvsMongoDb}, we compare the mean database access times by the Lambda function and the \texttt{Flask} web application to the \texttt{DyanamoDB} and \texttt{MongoDB} respectively, in five regions. Clearly, in all regions the access time of the database in serverless setup is significantly higher than the VM based setup. The most probable reason behind this is since in serverless the database and the Lambda function are not necessarily hosted on the same physical host, the \texttt{DynamoDB} access involves network latency, in contrast to \texttt{MongoDB} which is hosted in the same VM. Overall, the database access time in serverless architecture is nearly $14$ times of that in traditional VM based architecture if the database is hosted in the same VM.

In the second scenario, we compare the latency of the complex serverless architectures with varying critical path lengths. We take the smallest such setup is of length $1$ with only one Lambda function and a database (\texttt{DynamoDB}) just like the first case. For the architecture of critical path length $2$, the longest path involves two Lambda functions and one database at the end. Similarly, we deploy such setups of critical paths of length $3$, $4$, and $5$. Each of the Lambda function has negligible computation and their individual response times are less than $5ms$. This allows us to monitor only the overhead due to the chaining of serverless components through network calls. Figure \ref{fig:serverlessCascadingData.png} shows the box and whisker plot of the distribution of response time for 100 user requests. We can observe that the response time of the system increases steadily with increasing critical path length. The mean latency of the system increases by 7.6 times from $50ms$ in case of length 1 to $430ms$ for length 5. Thus, as the serverless architectures scale up, it incurs more and more latency overhead.

\section{Serverless Caching}\label{caching}
In our experiments, we observe that difference in latencies when using traditional approaches and while using serverless architecture is huge even when the cold starts are ignored. In both of the scenarios, the overall latency is a result of the network delay in between the components of the serverless architecture. Thus, in order to work around this latency and improve the overall response time of the system, we can introduce a cache so that recomputing/refetching previously computed results can be avoided.
Hence, we propose two different kinds of caching techniques and through our experiments, we try to determine how much improvement we can gain with caching.

Unlike in traditional cloud VMs, in Lambda functions, there is no provision to install an in-memory cache service such as \texttt{Memcache}, \texttt{Redis} etc. within the function. 
Instead, AWS provides separate external cache services such as \texttt{Redis} and \texttt{Memcache} through \texttt{Amazon ElastiCache}. Although using it will avoid computation/fetching time, these external caches are not hosted in the same container or host just like the other components. As a result, they also have the overhead of network calls.
In order to reduce the latency further by avoiding any network calls, we propose an in-memory internal cache by leveraging the Lambda container's memory persistency across multiple consecutive requests in short intervals and Javascript's asynchronous function calls for cache updates.

\begin{figure}[h]
	\center
	\includegraphics[width=0.8\linewidth]{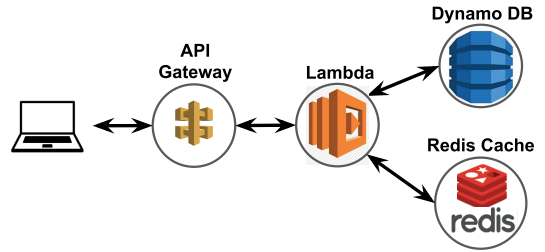}
	\caption{Redis as a cache}\label{fig:rediscache}
\end{figure}

\begin{figure}[h]
	\center
	\includegraphics[width=0.8\linewidth]{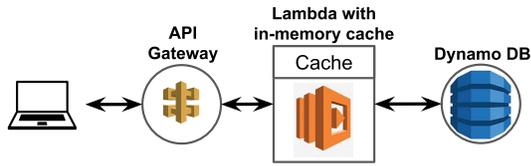}
	\caption{In-memory cache}\label{fig:memorycache}
\end{figure}

\textbf{Internal cache:} Here, by `internal', we mean internal to the runtime executing the Lambda function. AWS Lambda shares global variables in between different function calls of the same Lambda function. But, this feature comes with its own constraints. The global variables are shared between only those function calls which are sharing a session. A session starts with a cold start when the first request arrives and the function container is deployed. The subsequent requests are served by the same container. If there is a considerable time gap between the requests, the function container may be suspended and the session is closed. At this point, all the data stored in the global variables is also lost.
Therefore, if we use the memory of the function container for caching, whenever the frequency of requests drops, the container may be suspended which results in invalidation of the whole cache. So, we can say that in order to keep such a cache warm, the frequency of requests should not drop below a certain threshold.

We implement the in-memory cache in \texttt{Node.js} runtime as a global object variable. For read calls, we first check if the result of the request is already available in the memory. If it is available (cache hit), then the result is immediately returned as the response. In case of a miss, the result is computed/fetched and then saved in the cache before returning the response. Thus, if the cache is warm, the response time would be very low. But, if the cache is cold, the response time would be more as the time is spent in fetching the data from the database or other modules.

For write calls, in order to improve the latency, instead of synchronously updating the database, the function delegates the actual writing to another Lambda function asynchronously and spends time only in doing any preprocessing of the data to be written. This is implemented using Javascript's asynchronous function calls.

%


\section{Evaluation}\label{eval}
\begin{figure}[h]
\center
\includegraphics[width=0.9\linewidth]{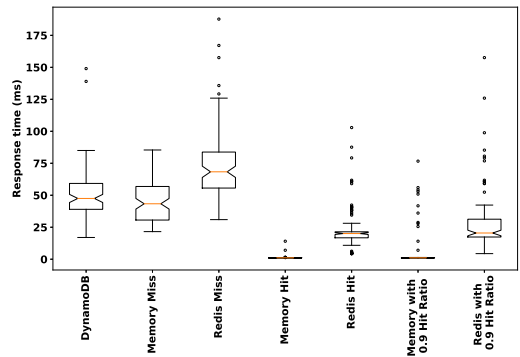}
		\caption{Comparison of response time using Redis cache, our proposed in-memory cache, and without cache.}\label{fig:RedisVsMemoryVsDynamoDB}
\end{figure}

In order to compare the performance with different caching techniques, we setup a simple database application involving \texttt{DynamoDB} and a Lambda function. Along with that, we implement \texttt{Redis} based cache of \texttt{Amazon ElastiCache} (Figure \ref{fig:rediscache}) and also our proposed in-memory internal caching (Figure \ref{fig:memorycache}). Figure \ref{fig:RedisVsMemoryVsDynamoDB} shows the distribution of the response time for different caching techniques for 100 user requests. We observe that using \texttt{Redis} cache improves the response time as compared to fetching data directly from \texttt{DynamoDB}. Moreover, our proposed in-memory cache further improves the response time significantly. With a hit ratio of 0.9, the in-memory internal cache reduces the response time by around 45ms. This is a significant improvement which is crucial for latency-sensitive realtime applications.

\section{Conclusion}\label{conlsn}
To rapidly develop and deploy complex services, serverless is a good candidate. However, the latency incurred between the different serverless components is an important issue, which may lead to degraded quality of experience for any service. In this work, we analyze different serverless architectures with AWS Lambda services and compare their performance in terms of latency with traditional VM based approach. Then, we compare some caching strategies which significantly improve the response time and ultimately the quality of experience of the end-users.  
An immediate extension of this work is to develop an in-memory caching system which remains warm in between Lambda sessions and consistent in case of multiple replicated function container instances.

\ifCLASSOPTIONcaptionsoff
  \newpage
\fi

\bibliographystyle{IEEEtran}
\bibliography{biblio}

\begin{thebibliography}{1}
\providecommand{\url}[1]{#1}
\csname url@samestyle\endcsname
\providecommand{\newblock}{\relax}
\providecommand{\bibinfo}[2]{#2}
\providecommand{\BIBentrySTDinterwordspacing}{\spaceskip=0pt\relax}
\providecommand{\BIBentryALTinterwordstretchfactor}{4}
\providecommand{\BIBentryALTinterwordspacing}{\spaceskip=\fontdimen2\font plus
\BIBentryALTinterwordstretchfactor\fontdimen3\font minus
  \fontdimen4\font\relax}
\providecommand{\BIBforeignlanguage}[2]{{%
\expandafter\ifx\csname l@#1\endcsname\relax
\typeout{** WARNING: IEEEtran.bst: No hyphenation pattern has been}%
\typeout{** loaded for the language `#1'. Using the pattern for}%
\typeout{** the default language instead.}%
\else
\language=\csname l@#1\endcsname
\fi
#2}}
\providecommand{\BIBdecl}{\relax}
\BIBdecl

\bibitem{Adzic:2017}
G.~Adzic and R.~Chatley, ``Serverless computing: Economic and architectural
  impact,'' in \emph{Proceedings of the 2017 11th Joint Meeting on Foundations
  of Software Engineering}, ser. ESEC/FSE 2017.\hskip 1em plus 0.5em minus
  0.4em\relax New York, NY, USA: ACM, 2017, pp. 884--889.

\bibitem{Wang_usenix_2018}
L.~Wang, M.~Li, Y.~Zhang, T.~Ristenpart, and M.~Swift, ``Peeking behind the
  curtains of serverless platforms,'' in \emph{Proceedings of the 2018 {USENIX}
  Annual Technical Conference ({USENIX} {ATC} 18)}.\hskip 1em plus 0.5em minus
  0.4em\relax Boston, MA: {USENIX} Association, Jul. 2018, pp. 133--146.

\bibitem{servelless_online_2019}
\BIBentryALTinterwordspacing
Selverless partners. [Online]. Available:
  \url{https://serverless.com/partners/}
\BIBentrySTDinterwordspacing

\bibitem{Lloyd_2018}
W.~Lloyd, M.~Vu, B.~Zhang, O.~David, and G.~H. Leavesley, ``Improving
  application migration to serverless computing platforms: Latency mitigation
  with keep-alive workloads,'' in \emph{Proceedings of the 2018 {IEEE/ACM}
  International Conference on Utility and Cloud Computing Companion, {UCC}
  Companion 2018, Zurich, Switzerland, December 17-20, 2018}, 2018, pp.
  195--200.

\bibitem{Bhattacharjee_USENIX_2019}
A.~Bhattacharjee, Y.~Barve, S.~Khare, S.~Bao, A.~Gokhale, and T.~Damiano,
  ``Stratum: A serverless framework for the lifecycle management of machine
  learning-based data analytics tasks,'' in \emph{Proceedings of the 2019
  {USENIX} Conference on Operational Machine Learning (OpML 19)}.\hskip 1em
  plus 0.5em minus 0.4em\relax Santa Clara, CA: {USENIX} Association, May 2019,
  pp. 59--61.

\bibitem{Akkus_Sand_Usenix_2018}
I.~E. Akkus, R.~Chen, I.~Rimac, M.~Stein, K.~Satzke, A.~Beck, P.~Aditya, and
  V.~Hilt, ``{SAND}: Towards high-performance serverless computing,'' in
  \emph{Proceedings of the 2018 {USENIX} Annual Technical Conference ({USENIX}
  {ATC} 18)}.\hskip 1em plus 0.5em minus 0.4em\relax Boston, MA: {USENIX}
  Association, Jul. 2018, pp. 923--935.

\bibitem{Oakes_USENIX_2018}
E.~Oakes, L.~Yang, D.~Zhou, K.~Houck, T.~Harter, A.~Arpaci-Dusseau, and
  R.~Arpaci-Dusseau, ``{SOCK}: Rapid task provisioning with
  serverless-optimized containers,'' in \emph{Proceedings of the 2018 {USENIX}
  Annual Technical Conference ({USENIX} {ATC} 18)}.\hskip 1em plus 0.5em minus
  0.4em\relax Boston, MA: {USENIX} Association, Jul. 2018, pp. 57--70.

\bibitem{Hong_USENIX_2018}
S.~Hong, A.~Srivastava, W.~Shambrook, and T.~Dumitras, ``Go serverless:
  Securing cloud via serverless design patterns,'' in \emph{Proceedings of the
  10th {USENIX} Workshop on Hot Topics in Cloud Computing (HotCloud 18)}.

\end{thebibliography}

\end{document}